\documentclass[twocolumn,pra,twocolumn,superscriptaddress]{revtex4}
\usepackage[latin9]{inputenc}
\usepackage{color}
\usepackage{amsmath}
\usepackage{amssymb}\usepackage{amsmath}
\usepackage{amsfonts}
\usepackage{amssymb}
\usepackage{graphicx}
\usepackage{tikz}
\usepackage{mathrsfs} 
\usepackage{float}
\usepackage{bbm}
\usepackage{epstopdf}
\usepackage{epsfig}
\usepackage{verbatim}
\usepackage{array}
\usepackage{setspace}
\usepackage{times}
\usepackage{tabularx}
\usepackage{setspace}
\usepackage{ulem}
\usepackage{soul}

\usepackage[unicode=true,
bookmarks=true,bookmarksnumbered=false,bookmarksopen=false,
breaklinks=false,pdfborder={0 0 1},backref=false,colorlinks=true]
{hyperref}
\hypersetup{
	linkcolor=magenta, urlcolor=blue, citecolor=blue, pdfstartview={FitH}, hyperfootnotes=false, unicode=true}

\newcolumntype{C}[1]{>{\centering\arraybackslash$}p{#1}<{$}}

\begin{document}

\title{Optimizing nonadiabatic geometric quantum gates against off-resonance error  in a silicon-based spin qubit}

\author{Liu-Jun Guo}
\affiliation{Guangxi Key Laboratory for Relativistic Astrophysics, School of Physical Science and Technology, Guangxi University, Nanning 530004, China}

\author{Hai Xu}
\affiliation{Guangxi Key Laboratory for Relativistic Astrophysics, School of Physical Science and Technology, Guangxi University, Nanning 530004, China}

\author{Zi-Yu Fang}
\affiliation{Guangxi Key Laboratory for Relativistic Astrophysics, School of Physical Science and Technology, Guangxi University, Nanning 530004, China}

\author{Tao Chen}
\affiliation{Guangdong Provincial Key Laboratory of Quantum Engineering and Quantum Materials, and School of Physics
	\\ and Telecommunication Engineering, South China Normal University, Guangzhou 510006, China}

\author{Kejin Wei}
\affiliation{Guangxi Key Laboratory for Relativistic Astrophysics, School of Physical Science and Technology, Guangxi University, Nanning 530004, China}

\author{Chengxian Zhang}\email{cxzhang@gxu.edu.cn}
\affiliation{Guangxi Key Laboratory for Relativistic Astrophysics, School of Physical Science and Technology, Guangxi University, Nanning 530004, China}

\date{\today}

\begin{abstract}

Geometric quantum gates are performed by using the geometric phase, making them particularly robust to the pulse amplitude error. However, in many systems, such as the silicon-based spin qubits, the off-resonance error is the dominant noise, which can cause dephasing and is always difficult to deal with for a geometric gate. Thus how to combat with the off-resonance error is very significant for the application of the geometric gates. A recent work in \emph{Phy. Rev. Appl. 16, 044005 (2021)} reveals that by inserting two  $\pi$-pulse into the evolution paths, the holonomic quantum gate is effective to suppress the pulse amplitude error, however it is still useless for combating the off-resonance error. Inspired by this work, we combine using the techniques of dynamical correction and path design. Surprisingly, we find that by picking up a specific evolution path inserted by only a  $\pi$-pulse, the obtained optimized geometric gate is robust to the off-resonance error, assuming the noise is static. Further, by calculating the filter function considering the realistic $1/f$-type noise in silicon, the related results show that the performance of the optimized geometric gate can also surpass both the conventional geometric gate and the naive dynamical gate constructed without using the geometric phase. Our results indicate dynamical correction is a powerful tool to improve the geometric quantum computation to achieve high-fidelity quantum gate in silicon.

\end{abstract}

\maketitle

\section{Introduction}

High-fidelity gate operation is crucial to realize fault-tolerant quantum computing and quantum task. Tremendous schemes such as composite quantum gates \cite{Wang.12,Bando.13, Zhang.17,Gevorgyan.21}, time optimal quantum gates \cite{Xiao.15,Geng.17, Dridi.20} and geometric gates \cite{Pachos.99,Zanardi.99,Duan.01,Zhu.02,Zhu.03,Zhao.17} have been proposed and have achieved great success over the past years. The geometric gate is standing out to realize high-fidelity quantum operations, since it uses the geometric phase rather than the traditional dynamical phase. Normally, there are two basic requirements \cite{Erik.15} for the construction of the geometric gates, namely the cancellation of the dynamical phase and the cyclic evolution in the parameter space. In this way, the implemented geometric gates can mitigate the fluctuation in the control Hamiltonian.

Geometric gates can be constructed by using either the Berry phase \cite{Berry.84}, or the Aharonov-Anandan (AA) phase \cite{Aharonov.87}, where they relate to the adiabatic and nonadiabatic evolution, respectively. In the early proposed schemes, many proposals are using the Berry phase. However, the adiabatic condition for the Berry phase requires overly long evolution time, making the gates suffering more decoherence. While the geometric gate based on the AA phase speeds up the evolution time, which is friendly for the experimental implementation. The AA-based universal geometric quantum gates have been realized in superconducting circuits \cite{abdumalikov.13,Tao.18,xu.18,Liu.19,egger.19,xu.20,Tao.20}, Nitrogen-vacancy center in diamond  \cite{Sekiguchi.17,Brian.17}, and trapped ions \cite{ai.20,ai.21}, etc. Typically, the conventional geometric gates based on orange-slice-shaped evolution loop \cite{Zhao.17} are particularly robust against the pulse amplitude error, i.e., the error in the Rabi frequency. But, the off-resonance error is always hard to deal with for the geometric gate \cite{Dong.21}, where this type of noise is the dominant noise source in many platforms. Therefore, how to combat the off-resonance error is important and significant for the implementation of geometric gates.

Silicon-based spin qubits in semiconductor quantum dot is promising to realize universal quantum computation due to its long coherence time and all-electric gate operation. Experimentally, the fidelity for both single- and the two-qubit gates have surpassed 99\% \cite{Noiri.22,madzik.22,xue.22, Adam.22}. Nevertheless, the implemented gate in silicon still suffers noises. Charge noise \cite{Bermeister.2014,Chan.17,Yoneda.18} and nuclear noise \cite{Huang.19} are assumed to be two main noise sources in silicon. Charge noise stems from the charge fluctuation near the quantum dot. While the nuclear noise is attributed to the residual $^{29}\rm{Si}$ nuclei. Experimentally, the single-qubit gate can be resonantly driven by applying an oscillating magnetic field, which can be generated by either using the electron spin resonance (ESR) or electric dipole spin resonance (EDSR). For the ESR-driven spin qubits, the dephasing is mostly owing to the residual $^{29}\rm{Si}$ nuclei \cite{Huang.19}. While for the EDSR-driven case, the charge noise can induce spin dephasing via the longitudinal spin-electric coupling \cite{Yoneda.18}. For both cases, the noises can induce the resonance frequency fluctuations and further the qubit dephasing. Therefore, we term them as the off-resonance errors in this work for simplicity. On the other hand, the valley-spin coupling \cite{Jock.22} would lead to unwanted relaxation for the spin qubits. Fortunately, a recent experimental work \cite{Ciriano.21} found that, by using the isotopic purification technique, the spin qubits can be operated with relaxation time as long as 9 s . In this work, we focus on the off-resonance error suppression by geometric gates and assume the evolution of the gate is unitary due to the long relaxation time. The relaxation effect on the gate fidelity is discussed in Appendix.~\ref{appx2}. The valley effect remains to be further studied.

Recently, it has been found that by using the dynamical corrected technique, the holonomic quantum gates can be substantially robust to the pulse amplitude error to fourth order \cite{Li.21}. In this protocol, to cancel out the dynamical phase accumulation, two decoupling $\pi$-pulse are symmetrically inserted into the middle point of the evolution paths, which add extra 2$\pi$ rotation, compared to the conventional holonomic quantum gate. In combination with the decoherence free subspace encoding \cite{Duan.97,Zanardi.97,Lidar.98} the collective dephasing noise can be further canceled out, since the noise performs certain symmetric in this subspace. Later, it was found that this protocol can be successfully extended to the abelian geometric gates \cite{Liang.22}. However, in both cases it is useless to suppress the off-resonance error, as seen in Appendix.~\ref{appx}, which is the targeted issue in this work.  Inspired by these works, in this work, we study the key issue of whether the geometric gate can suppress the off-resonance error by dynamical correction. By picking up a specific evolution path and inserting only a  $\pi$-pulse into the middle of this path, one is able to achieve an optimized geometric gate, which is equivalent to the conventional geometric gate. We surprisingly find that, the designed optimized geometric gate can be effectively robust to the off-resonance error. To verify the performance of the optimized geometric gate, we perform randomized benchmarking \cite{Emerson.05,Knill.08,Easwar.12,Zhang.17} and calculate the filter function \cite{Green.12,Green.13,Silva.14,Barnes.16}, considering the realistic experimental noise level. The results have clearly shown that the performance of the optimized geometric gate are superior to both the conventional geometric gate \cite{Zhao.17} and the naive dynamical gates that are constructed without using the geometric phase. Our results indicate that the geometric gate might benefit from dynamical correction. We emphasize that our method can not only be useful for the single-qubit gates in silicon but can also be easily extended to other systems like superconducting qubits with microwave-driven Hamiltonian as shown in Eq.~\ref{Hc}. In addition, it is also applied for the two-qubit case when the Hamiltonian in the related subspace has the similar form \cite{Tao.18}.

\section{Conventional geometric gates}\label{sec:con}

\begin{figure}
	\includegraphics[width=0.8\columnwidth]{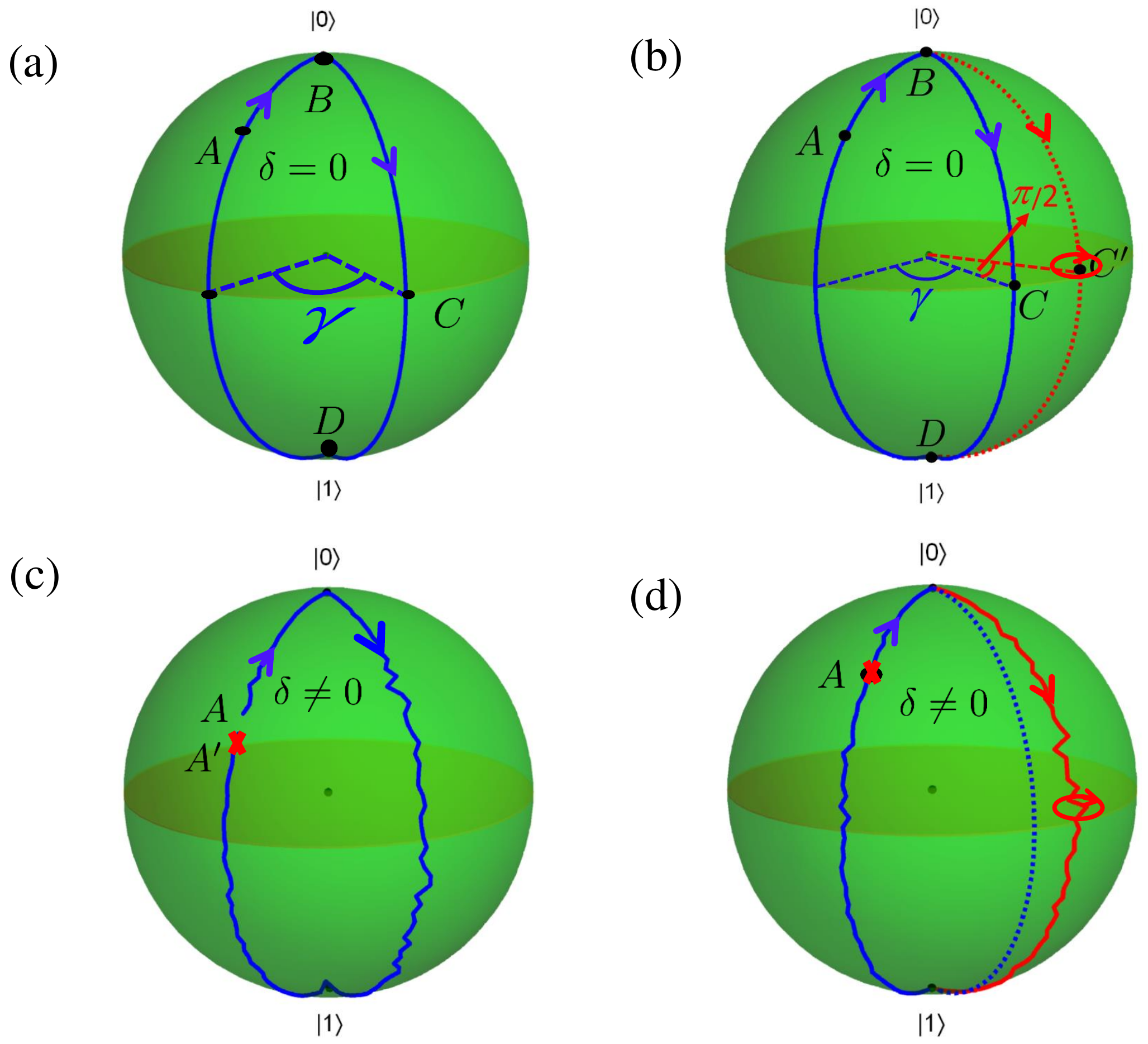}
	\caption{Schematic of the conventional and optimized geometric gates. (a) and (b) show the case without the off-resonance error, i.e., $\delta=0$. For the conventional geometric gate in (a), the dressed state evolves along the longitude A-B-C-D-A to fulfill a cyclic evolution. In (c), when the off-resonance error is existing, i.e., $\delta\neq0$, the dressed state cannot go back to the desired starting point A. While for the optimized geometric gate in (b), the dressed states evolves along another path A-B-$\rm{C'}$-D-A, where a $\pi$-pulse has been inserted into point $\rm{C'}$. In (d), the dressed state can still enclose a closed loop even though $\delta\neq0$.}
	\label{fig:qdot_bloch}
\end{figure}

Here, we present the approach to construct the conventional orange-slice-shaped geometric gates \cite{Zhao.17}, as shown in Fig.~\ref{fig:qdot_bloch}(a). This conventional geometric gate is based on a two-level system driven by a microwave. By setting $\hbar=1$, the control Hamiltonian reads 
\begin{equation}
\begin{aligned}
H_{c}(t)=&\frac{\Omega(t)}{2}(\cos\varphi(t)\ \sigma_{x}+\sin\varphi(t)\ \sigma_{y}).
\end{aligned}
\label{Hc}
\end{equation}
Here, the Pauli matrix is written in the computational basis  $|0\rangle$ and $|1 \rangle$. $\Omega(t)$ and $\varphi(t)$ are the time-dependent amplitude (Rabi frequency) and phase of the microwave field, respectively. To obtain the pure geometric gates, we divide the whole evolution time of any geometric gate into three segments. The two parameters during each segment should satisfy
\begin{equation}
\begin{aligned}
\int_{0}^{T_1}\Omega(t)dt&= \theta, \quad \left\{\varphi(t)\equiv\phi-\frac{\pi}{2},t\in\left[0,T_1\right]\right\}\\
\int_{T_1}^{T_2}\Omega(t)dt&= \pi, \quad \left\{\varphi(t)\equiv\phi+\gamma+\frac{\pi}{2},t\in\left[T_1,T_2\right]\right\}\\
\int_{T_2}^{T}\Omega(t)dt&= \pi-\theta, \quad \left\{\varphi(t)\equiv\phi-\frac{\pi}{2},t\in\left[T_2,T\right]\right\}.\\
\end{aligned}
\label{eq:seg}
\end{equation}
At the final evolution time $T$, which is determined by $\int_{0}^{T}\Omega(t)dt=2\pi$, the total evolution operator is 
\begin{equation}
\begin{aligned}
U_{g}(T)&=U(T,T_2)U(T_2,T_1)U(T_1,0)\\
&=\cos\gamma\hat{I}-i\sin\gamma \left(\begin{array}{cc}\cos\theta&\sin\theta e^{-i \phi}\\ \sin\theta e^{i \phi}&-\cos\theta\end{array}\right)\\
&=e^{i\gamma\vec{n}\cdot \vec{\sigma}},
\end{aligned}
\label{eq:U_gs}
\end{equation}
where $\vec{n}=\left(\sin\theta\cos\phi,\sin\theta\sin\phi,\cos\theta\right)$ can be regarded as the unit vector on the Bloch sphere.
In this way, $U_{g}(T)$ can implement arbitrary single-qubit rotation on the Bloch sphere, namely, it is universal. 

Generally, a geometric gate needs to satisfy the cyclic evolution and parallel-transport conditions \cite{Erik.15}. To demonstrate $U_{g}(T)$ is being a pure geometric gate, we introduce a pair of dressed states
\begin{equation}
\begin{aligned}
|\psi_{+}\rangle&=\cos \frac{\theta}{2}|0\rangle+\sin \frac{\theta}{2} e^{i \phi}|1\rangle,\\
|\psi_{-}\rangle&=\sin \frac{\theta}{2} e^{-i \phi}|0\rangle-\cos \frac{\theta}{2}|1\rangle.
\end{aligned}
\label{eq:dress}
\end{equation}
In the dressed-states representation, $U_{g}(T)$ can be rewritten as 
\begin{equation}
\begin{aligned}U_{g}(T)=e^{ {i}\gamma}|\psi_{+}\rangle\langle\psi_{+}| +e^{- {i} \gamma} |\psi_{-}\rangle\langle\psi_{-}|
\end{aligned}.
\label{eq:U_gdress}
\end{equation}
After the action of the operator, the dressed states become
\begin{equation}
\begin{aligned}
\left|\psi_{+}(T)\right\rangle&=U_{g}(T)\left|\psi_{+}\right\rangle=e^{i \gamma}\left|\psi_{+}\right\rangle, \\
\left|\psi_{-}(T)\right\rangle&=U_{g}(T)\left|\psi_{-}\right\rangle=e^{-i \gamma}\left|\psi_{-}\right\rangle.
\end{aligned}
\label{eq:cyclic}
\end{equation}
It is clear that $|\psi_{+}(t)\rangle$ ($|\psi_{-}(t)\rangle$) has fulfilled a cyclic evolution at the final evolution time $T$, and obtained the corresponding global phase factor $\gamma$ ($-\gamma$). The evolution path of $|\psi_{+}(t)\rangle$ is visualized in Fig.~\ref{fig:qdot_bloch}(a). Specifically, it travels always along the longitude of the Bloch sphere, denoted as path A-B-C-D-A, and encloses a orange-slice-shaped loop at the final time. On the other hand, the parallel-transport condition can be easily verified via 
\begin{equation}
\begin{aligned}
\left\langle\psi_{\pm}(0)\left|U_{g}^{\dagger}(t) H_{c}(t) U_{g}(t)\right| \psi_{\pm}(0)\right\rangle=0.
\end{aligned}
\label{eq:parallel}
\end{equation}
Therefore, the accumulated dynamical phase throughout the evolution is zero and the acquired global phase factor $\gamma$ represents the desired pure geometric phase.

Below we discuss the robustness of the conventional geometric gates by comparing its fidelity with the naive dynamical gate, which is constructed without using the geometric phase. We begin by introducing how to implement the dynamical gates. For the control Hamiltonian in Eq.~\ref{Hc}, by holding $\Omega(t)\equiv\Omega$ and $\varphi(t)\equiv\varphi$ for a time $\chi/\Omega$, a one-piece of dynamical gate is obtained as
\begin{equation}
\begin{aligned}
R(\hat{r}, \chi)&\equiv \exp\left[-i\frac{\chi}{2}(\cos\varphi\ \sigma_{x}+\sin\varphi\ \sigma_{y})\right],
\end{aligned}
\label{eq:Rot}
\end{equation}
where $R(\hat{r}, \chi)$ denotes arbitrary rotation by an angle $\chi$ in the $x$-$y$ plane. For other dynamical rotations that are out of this plane, they can be decomposed into the ${x}$-${y}$-${x}$ composite pulse sequence \cite{Wang.14}
\begin{equation}
\begin{aligned}
R(\hat{x},\chi_c)R(\hat{y},\chi_b)R(\hat{x},\chi_a),
\end{aligned}
\label{eq:dynamical}
\end{equation}
where the values of $\chi_{i}$ ($i=a,b,c$) is determined by the specific rotation. It is clear that an $x$-axis or $y$-axis rotation is the elementary unit. Note that one should not confuse this composite pulse sequence with the geometric gate as described in Eq.~\ref{eq:U_gs}, since it usually cannot satisfy the cyclic evolution and parallel-transport conditions for the geometric gate.

\begin{figure*}
	\includegraphics[width=1.8\columnwidth]{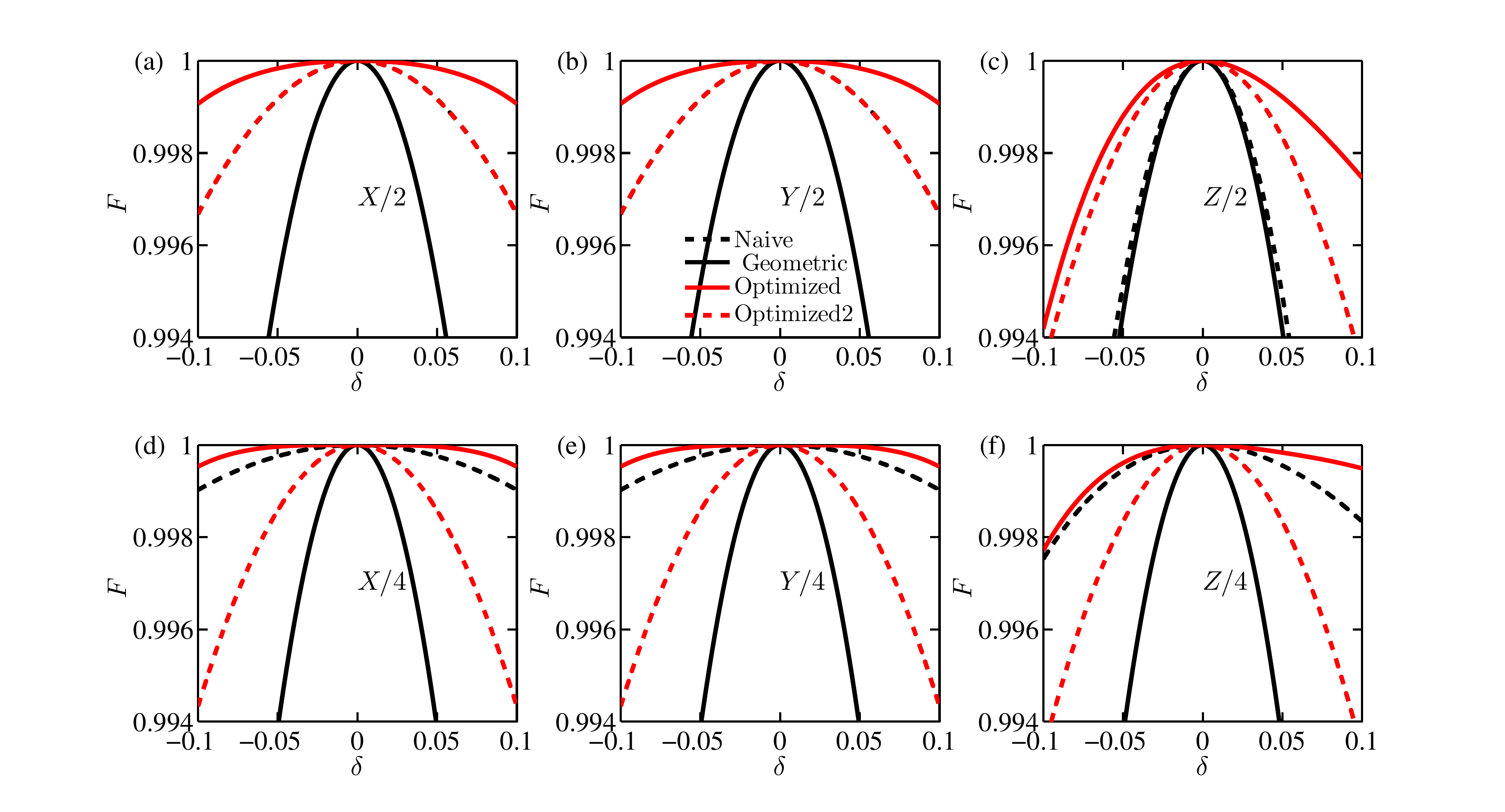}
	\caption{Robustness of the optimized geometric gates (denoted as red solid lines) compared with the conventional geometric gates (denoted as black solid lines) and the naive dynamical gates (denoted as black dashed lines). The red dashed line denotes the optimized geometric gate considering the  $\pi$-pulse is perfect.}
	\label{fig:robust}
\end{figure*}

In the implementation of the quantum operation, both the geometric and dynamical quantum gates can suffer mainly two types of errors, and the control Hamiltonian becomes $H_{c}'(t)=\frac{(1+\epsilon)\Omega(t)}{2}(\cos\varphi(t) \sigma_{x}+\sin\varphi(t) \sigma_{y})+\frac{\Omega(t)\delta}{2} \sigma_{z}$. $\epsilon$ relates to the fluctuation for the Rabi frequency due to the imprecise pulse control of the microwave field. We term it as the pulse amplitude error. While $\delta$ is the off-resonance error, which relates to the fluctuation transverse to the microwave field leading to dephasing. In the experimental environment for silicon-based spin qubits, both the charge noise \cite{Yoneda.18} and the residual ${ }^{29} \mathrm{Si}$ nuclei \cite{Huang.19} can cause this off-resonance error. The robustness of the conventional geometric gates against these two types of errors can be analytically derived by performing a Taylor series expansion of the fidelity. Here we take an arbitrary $x$-axis rotation by an angle $\chi$ as the example, since it is the elementary gate as seen in Eq.~\ref{eq:dynamical}:
\begin{equation}
\begin{aligned}
\mathcal{F}_{d}^{\delta}(\hat{x},\chi)&\simeq 1+\frac{1}{4}(\cos\chi-1)\delta^{2},\\ 
\mathcal{F}_{g}^{\delta}(\hat{x},\chi)&\simeq 1-2\cos^{4}\frac{\chi}{4}\delta^{2},\\
\mathcal{F}_{d}^{\epsilon}(\hat{x},\chi)&\simeq 1-\frac{\chi^{2}}{8}\epsilon^{2},\\ 
\mathcal{F}_{g}^{\epsilon}(\hat{x},\chi)&\simeq 1-\frac{\pi^{2}}{2}\sin^{4}\frac{\chi}{4}\epsilon^{2}.
\end{aligned}
\label{eq:expand}
\end{equation}
Here,  $\mathcal{F}_{d}(\hat{x},\chi)$ and $\mathcal{F}_{g}(\hat{x},\chi)$ denote the fidelity for the dynamical and conventional geometric gate, respectively. Note that here we have assumed $|\epsilon |,\ |\delta | \ll \Omega$, thus they can be treated as perturbations. Meanwhile, we have treated them as constant values and the time-dependent property will be discussed later. It is easy to find that in the region $-\pi\leq \chi\leq \pi$, $\mathcal{F}_{d}^{\delta}(\hat{x},\chi)\geq \mathcal{F}_{g}^{\delta}(\hat{x},\chi)$ and $\mathcal{F}_{d}^{\epsilon}(\hat{x},\chi)\leq \mathcal{F}_{g}^{\epsilon}(\hat{x},\chi)$, namely, the conventional geometric gate is robust to the pulse amplitude error while sensitive to the off-resonance error. The case for the $y$-axis rotation is similar, thus we would not show the result again. As stated above, the off-resonance error is the dominant noise for a silicon-based spin qubits. Therefore, the key to realizing high-fidelity quantum operation for a spin qubit in silicon is to seek for approaches to combat the off-resonance error. Hereafter we leave alone the pulse amplitude error and focus on the off-resonance error. In Fig.~\ref{fig:robust}, we compare the robustness between the dynamical gates and the conventional geometric gates.

\section{Optimized geometric quantum gates by dynamical correction}\label{sec:comparison}
In order to combat the off-resonance error, we combine the techniques of dynamical correction and path design. As shown in Fig.~\ref{fig:qdot_bloch}(b), we carefully choose a new evolution path (denoted as the red dashed line) and insert a corrected $\pi$-pulse into this path accordingly, such that the total evolution path is divided into five pieces. Specifically, $\Omega(t)$ and $\varphi(t)$ during each part satisfy 
\begin{equation}
\begin{aligned}
\int_{0}^{T_1}\Omega(t)dt&= \theta, \quad \left\{\varphi(t)\equiv\phi-\frac{\pi}{2},t\in\left[0,T_1\right]\right\}\\
\int_{T_1}^{T_2}\Omega(t)dt&= \pi/2, \quad \left\{\varphi(t)\equiv\phi+\gamma+\pi,t\in\left[T_1,T_2\right]\right\}\\
\int_{T_2}^{T_3}\Omega(t)dt&= \pi, \quad \left\{\varphi(t)\equiv\phi+\gamma+\frac{3\pi}{2},t\in\left[T_2,T_3\right]\right\}\\
\int_{T_3}^{T_4}\Omega(t)dt&= \pi/2, \quad \left\{\varphi(t)\equiv\phi+\gamma+\pi,t\in\left[T_3,T_4\right]\right\}\\
\int_{T_4}^{T}\Omega(t)dt&= \pi-\theta, \quad \left\{\varphi(t)\equiv\phi-\frac{\pi}{2},t\in\left[T_4,T\right]\right\}.
\end{aligned}
\label{eq:seg2}
\end{equation}
At the end of the evolution time, namely $\int_{0}^{T}\Omega(t)dt=3\pi$, the resulted operator is
\begin{equation}
\begin{aligned}
U_{g,opt}(T)&=U(T,T_4)U(T_4,T_3)U(T_3,T_2)U(T_2,T_1)U(T_1,0)\\
&=e^{i\gamma\vec{n}\cdot \vec{\sigma}},
\end{aligned}
\label{eq:U_gs2}
\end{equation}
which is the same as the conventional geometric gate in Eq.~\ref{eq:U_gs}. Hereafter, we term it as the optimized geometric gate.

Comparing the conventional geometric gate and the optimized one in Fig.~\ref{fig:qdot_bloch}(a) and \ref{fig:qdot_bloch}(b), the evolution paths for the first and the last parts are the same. While the differences are related to the second part, and there are two main differences between them. First, the longitude denoted as B-C-D for the conventional geometric gate has been rotated by an angle $\pi/2$, which turns to be another longitude B-$\rm{C'}$-D. Second, this new path has been further divided into three segments, where a corrected $\pi$-pulse has been inserted into the middle point of the new longitude. As shown in Fig.~\ref{fig:robust}, comparing to the dynamical and the conventional geometric gates, the optimized geometric gates are more insensitive to the off-resonance error. By further performing a Taylor series expansion, one finds
\begin{equation}
\begin{aligned}
\mathcal{F}_{g,opt}^{\delta}(\hat{x},\chi)&\simeq 1-2\sin^{4}\frac{\chi}{4}\delta^{2}.
\end{aligned}
\label{eq:expand2}
\end{equation}
It is clear that $\mathcal{F}_{g,opt}^{\delta}(\hat{x},\chi)\geq \mathcal{F}_{d}^{\delta}(\hat{x},\chi),\mathcal{F}_{g}^{\delta}(\hat{x},\chi)$ in the considered region $-\pi\leq \chi\leq \pi$. Therefore the robustness of the optimized geometric gate against the off-resonance error is verified. The case for the $y$-axis rotation is similar. We note that the evolution time for the optimized geometric gate is prolonged, however, the relaxation effect on the qubit can still be safely ignored considering the rather long coherence time in silicon. The detail can be seen in Appendix.~\ref{appx2}.

\begin{figure}
	\includegraphics[width=0.9\columnwidth]{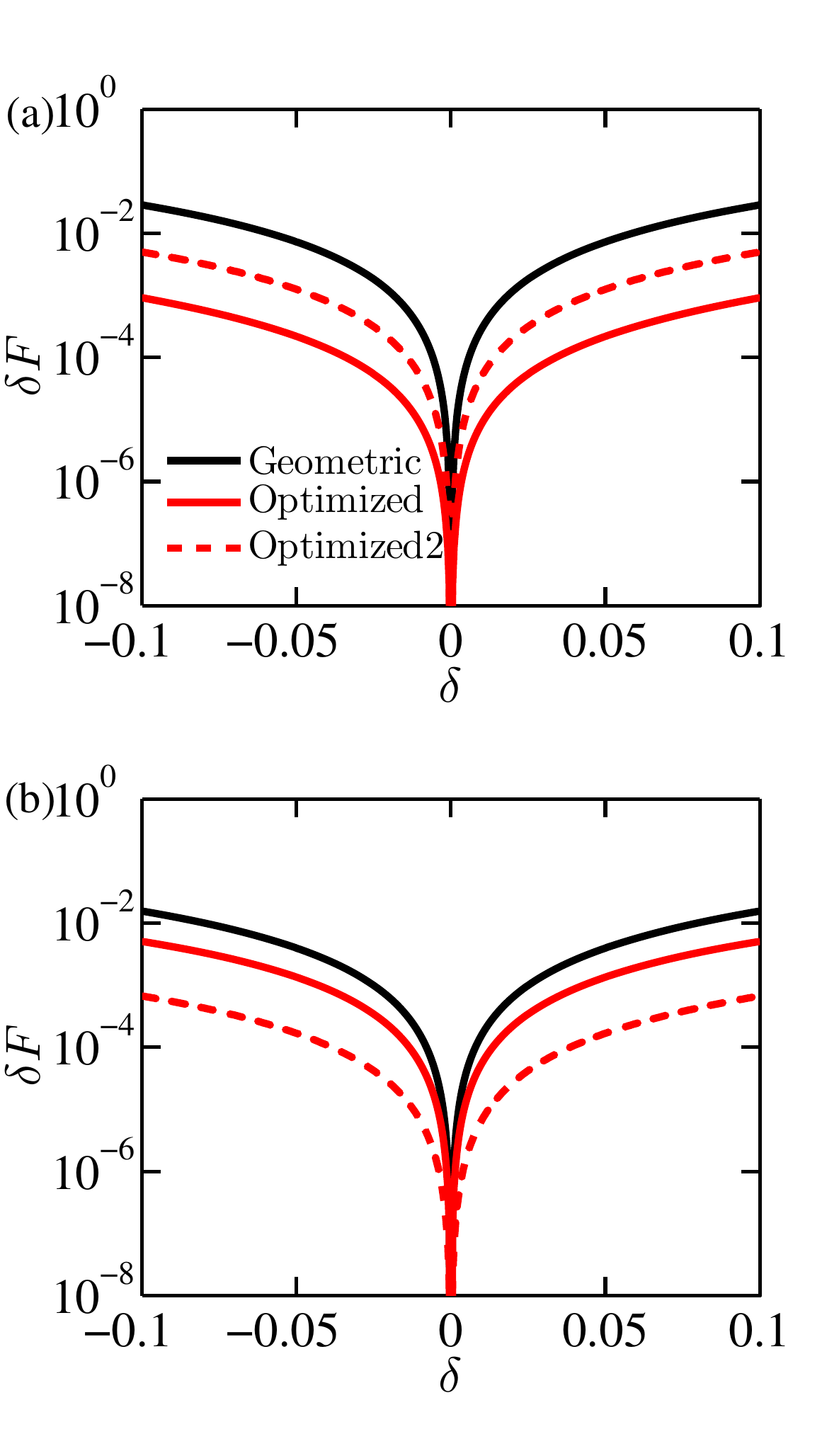}
	\caption{Infidelity for the $x$-axis rotations by an angle $\chi$, which are defined as $\delta F=1-|\langle \psi_{+}(0)|\psi_{+}^{\delta}(T)\rangle|^{2}$. The common parameters are $\theta=\pi/2$ and $\phi=0$. The rotation angles in (a) and (b) are $\chi=\pi/2$ and $5\pi/6$, respectively.}
	\label{fig:inf}
\end{figure}

Then, we explain why the designed optimized geometric gates can improve the robustness of the conventional geometric gates. The key to realizing a robust geometric gate is to enclose a closed loop even though the off-resonance error is existing. As shown in Fig.~\ref{fig:qdot_bloch}(c), when the off-resonance error is existing, the dressed state $|\psi_{+}\rangle$ ($|\psi_{-}\rangle$) cannot go back to the starting point at the final evolution time, namely, $|\psi_{+}(0)\rangle\neq |\psi_{+}^{\delta}(T)\rangle$ ($|\psi_{-}(0)\rangle\neq |\psi_{-}^{\delta}(T)\rangle$). According to Eq.~\ref{eq:U_gdress}, the evolution operator turns to be 
\begin{equation}
	\begin{aligned}U_{g,\delta}(T)=e^{ {i}\gamma^{\delta}}|\psi_{+}^{\delta}(T)\rangle\langle\psi_{+}(0)| +e^{- {i} \gamma^{\delta}} |\psi_{-}^{\delta}(T)\rangle\langle\psi_{-}(0)|.
	\end{aligned}
	\label{eq:U_gdress2}
\end{equation}
For comparison, in Fig.~\ref{fig:qdot_bloch}(d), the dressed state can still enclose a closed loop even though $\delta\neq0$ due to the inserted  $\pi$-pulse and the chosen different evolution path denoted as B-$\rm{C'}$-D. To quantitatively study the deviation between the initial and  the final dressed state, we define the infidelity as $\delta F=1-|\langle \psi_{+}(0)|\psi_{+}^{\delta}(T)\rangle|^{2}$. In Fig.~\ref{fig:inf}, we show two typical results for the $x$-axis rotations to verify the superiority of the optimized geometric gates. It is shown that the infidelity for the optimized geometric gates are always lower than the one for the conventional geometric gates, which means the inserted  $\pi$-pulse can effectively help to enclose a closed loop. Ideally, the inserted $\pi$-pulse is expected to be perfect, namely, the off-resonance is zero. We surprising find that even though the inserted $\pi$-pulse is imperfect, the performance for the imperfect geometric gates can sometimes outperform the perfect one, as seen in Fig.~\ref{fig:inf}(a) and Fig.~\ref{fig:robust}. Also, we have observed the case, where the perfect $\pi$-pulse performs better, as seen in Fig.~\ref{fig:inf}(b). We therefore conclude that whether the perfect $\pi$-pulse can offer external improvement sensitively depends on the specific rotations. Averagely, whether the $\pi$-pulse is perfect or not would not affect the averaged fidelity, which can be seen from the benchmarking results (see Sec.~\ref{sec:rb} below) as shown in Fig.~\ref{fig:strb} and Fig.~\ref{fig:rb2}.

\section{simulation on single-qubit gates in silicon}\label{sec:rb}


\begin{figure}
	\includegraphics[width=0.92\columnwidth]{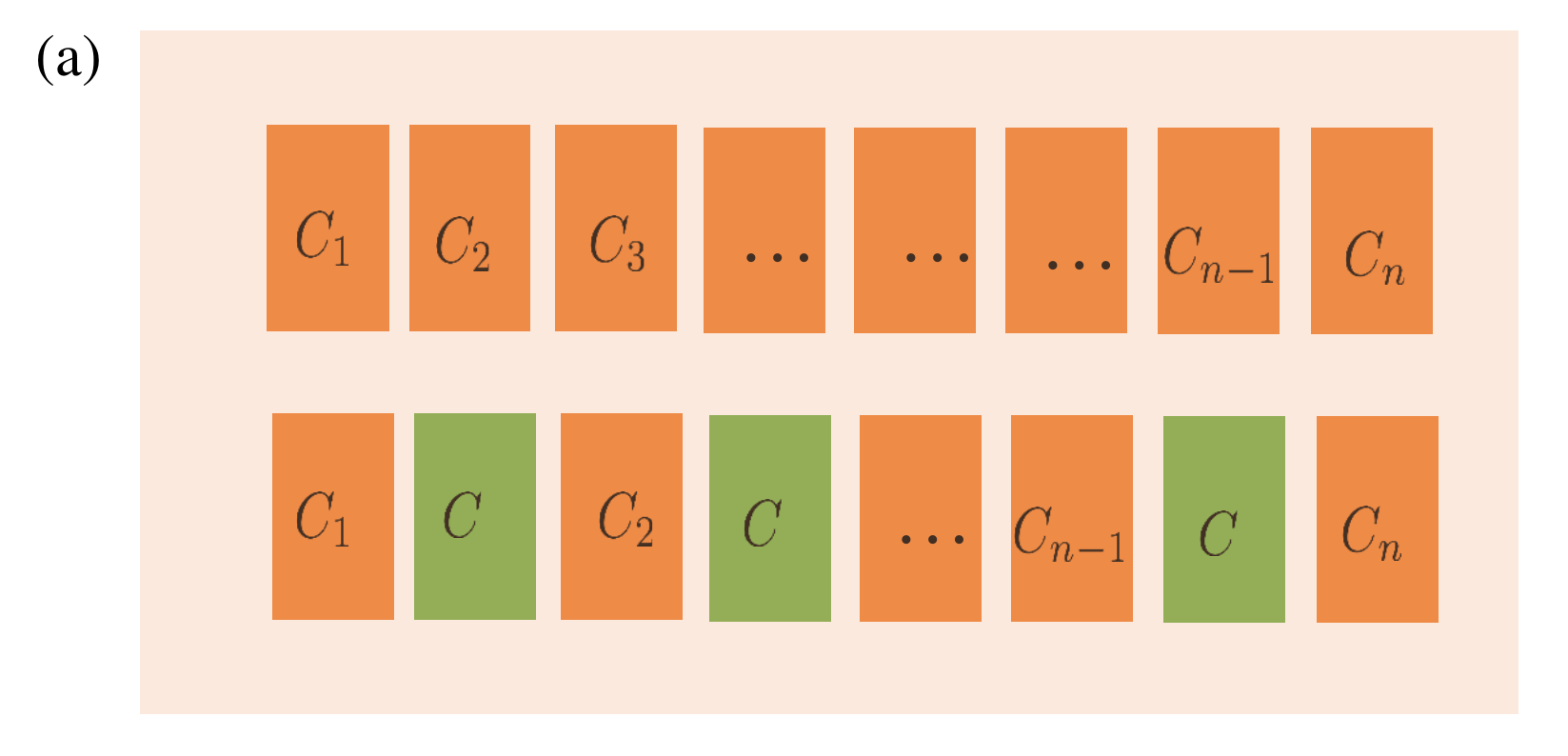}
	\includegraphics[width=0.93\columnwidth]{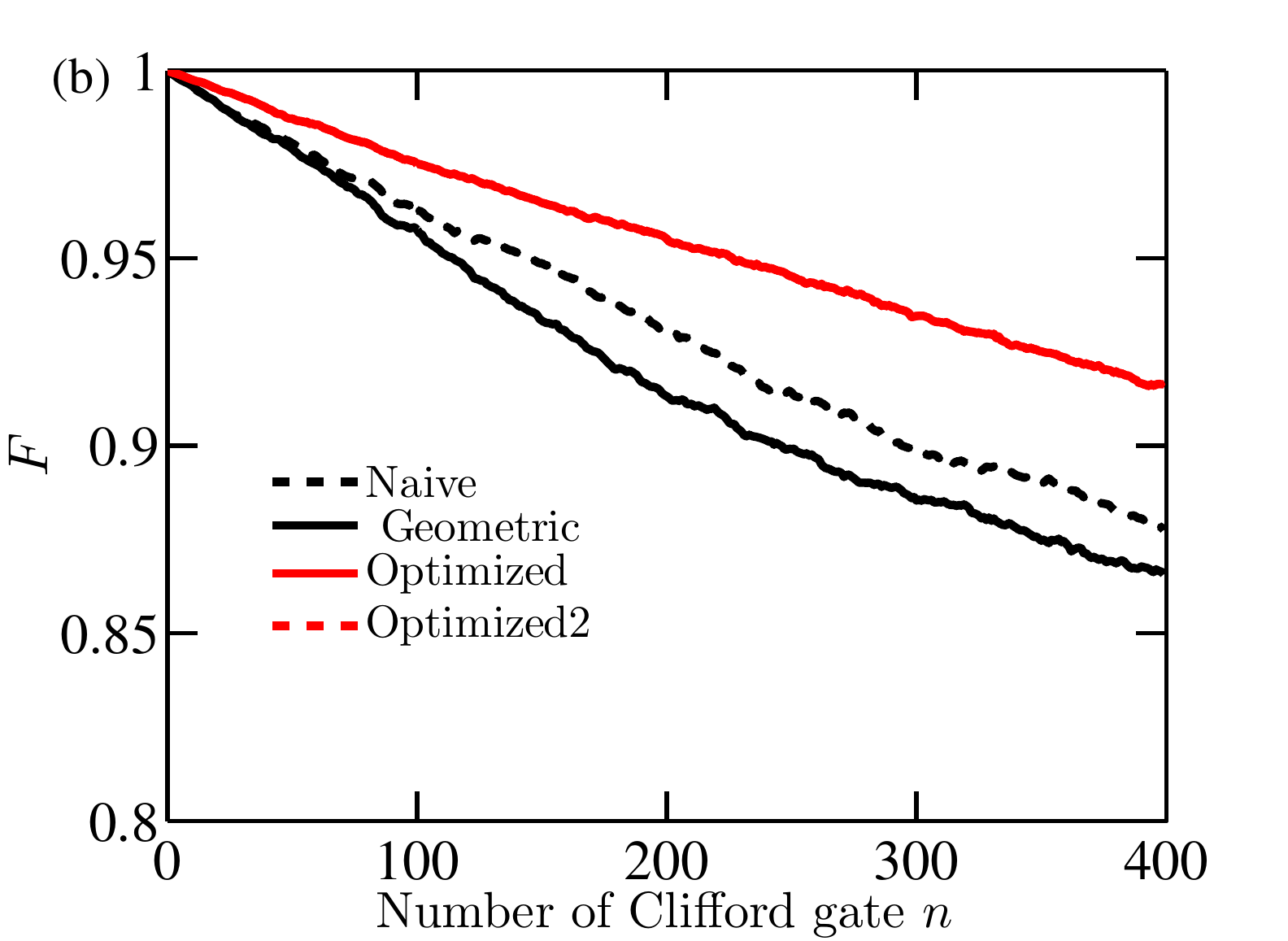}
	\caption{Randomized benchmarking results. (a) Schematic of the standard and the interleaved randomized benchmarking. The gate denoted as $C_{i},\ i=1,2,...,n$ is randomly drawn from the Clifford group, where $n$ is the number of the used Clifford gates. $C$ denotes the target gate for interleaving. (b) Standard randomized benchmarking results. The noise is assumed to be static with Gaussian distribution, i.e., $\sigma_{\delta}^{2}: \mathcal{N}\left(0, \sigma_{\delta}^{2}\right)$, where $\sigma_{\delta}$ is the variance of the noise. In the simulation, we take $\sigma_{\delta}/\Omega=0.02$ \cite{Huang.19,Zhang.20}. }
	\label{fig:strb}
\end{figure}

\begin{figure*}
	\includegraphics[width=1.75\columnwidth]{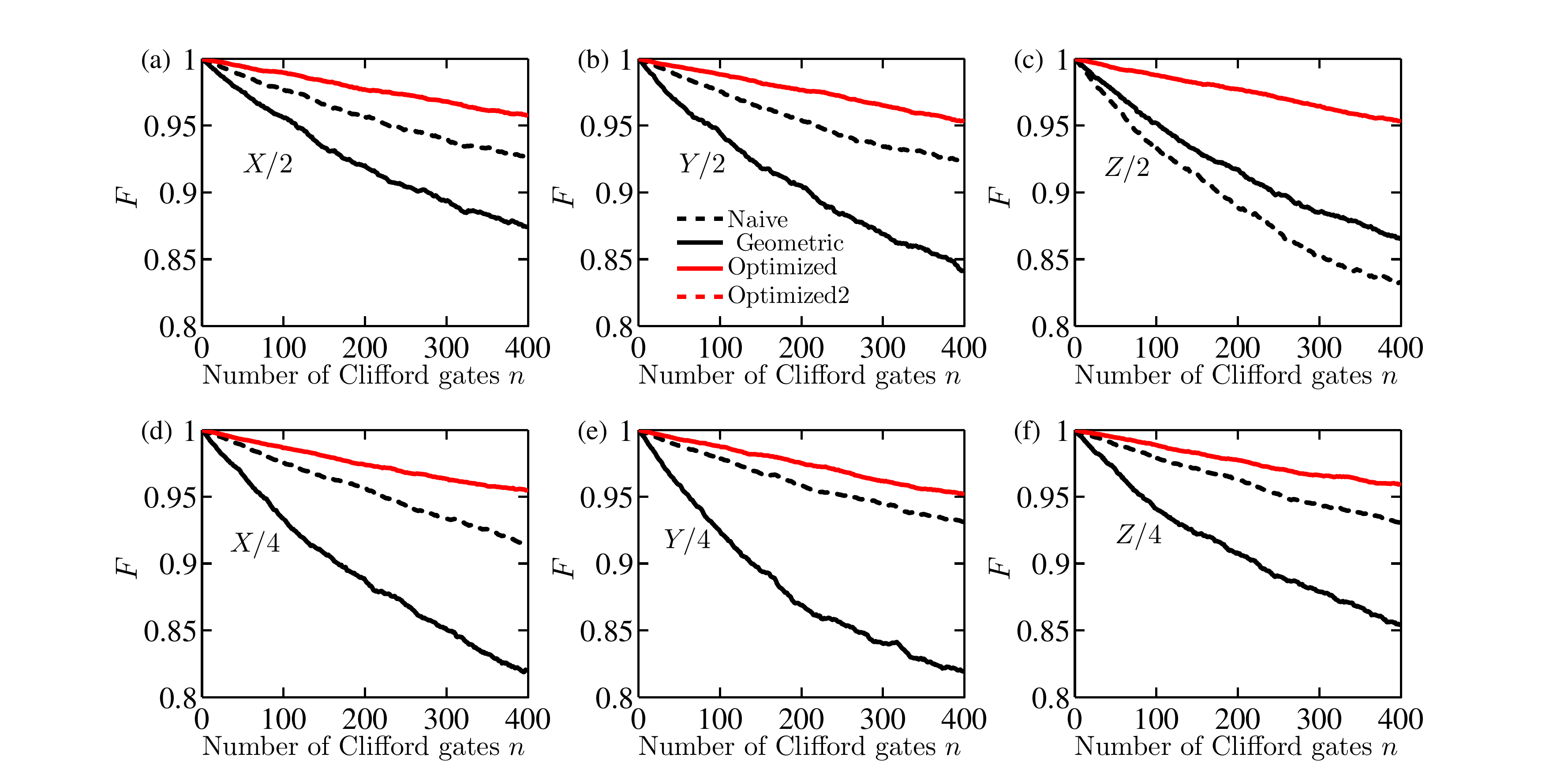}
	\caption{Interleaved randomized benchmarking results. The noise is assumed to be static with Gaussian distribution, i.e., $\sigma_{\delta}^{2}: \mathcal{N}\left(0, \sigma_{\delta}^{2}\right)$, where $\sigma_{\delta}$ is the variance of the noise. In the simulation, we take $\sigma_{\delta}/\Omega=0.02$ \cite{Huang.19,Zhang.20} .}
	\label{fig:rb2}
\end{figure*}

The superiority of the optimized geometric gate can be further verified by performing randomized benchmarking \cite{Emerson.05,Knill.08,Easwar.12,Zhang.17}, the schematic of which is shown in Fig.~\ref{fig:strb}(a). Normally, there are two types of randomized benchmarking, namely, the standard and the interleaved randomized benchmarking. For the standard randomized benchmarking, one can get the averaged fidelity over the given gate sequences and noise spectrum. The gate sequences are randomly drawn from the Clifford group, which is composed of 24 specific single-qubit gates \cite{Zhang.17}. While the interleaved randomized benchmarking \cite{Easwar.12} is the variant of the standard one, where the gate sequence is interleaved with specific target gate, such that one can also obtain the benchmarking result for this gate . In the randomized benchmarking simulation, we consider the static noise model, (the time-dependent effect will be considered later), and assume the noise is with Gaussian distribution, i.e., $\sigma_{\delta}^{2}: \mathcal{N}\left(0, \sigma_{\delta}^{2}\right)$, where $\sigma_{\delta}$ is the variance of the noise. According to Ref.~\cite{Huang.19,Zhang.20}, we assume $\sigma_{\delta}/\Omega=0.02$. The standard randomized benchmarking results are shown in Fig.~\ref{fig:strb}(b). By fitting the fidelity curves to the expression $\left(1+e^{-d n}\right) / 2$ \cite{Wang.2016}, one is able to obtain the averaged error per gate $d$. Here, $n$ is the number of the Clifford gates. Therefore, the averaged fidelity is $\mathcal{F}=1-d$.

\begin{figure*}
	\includegraphics[width=1.8\columnwidth]{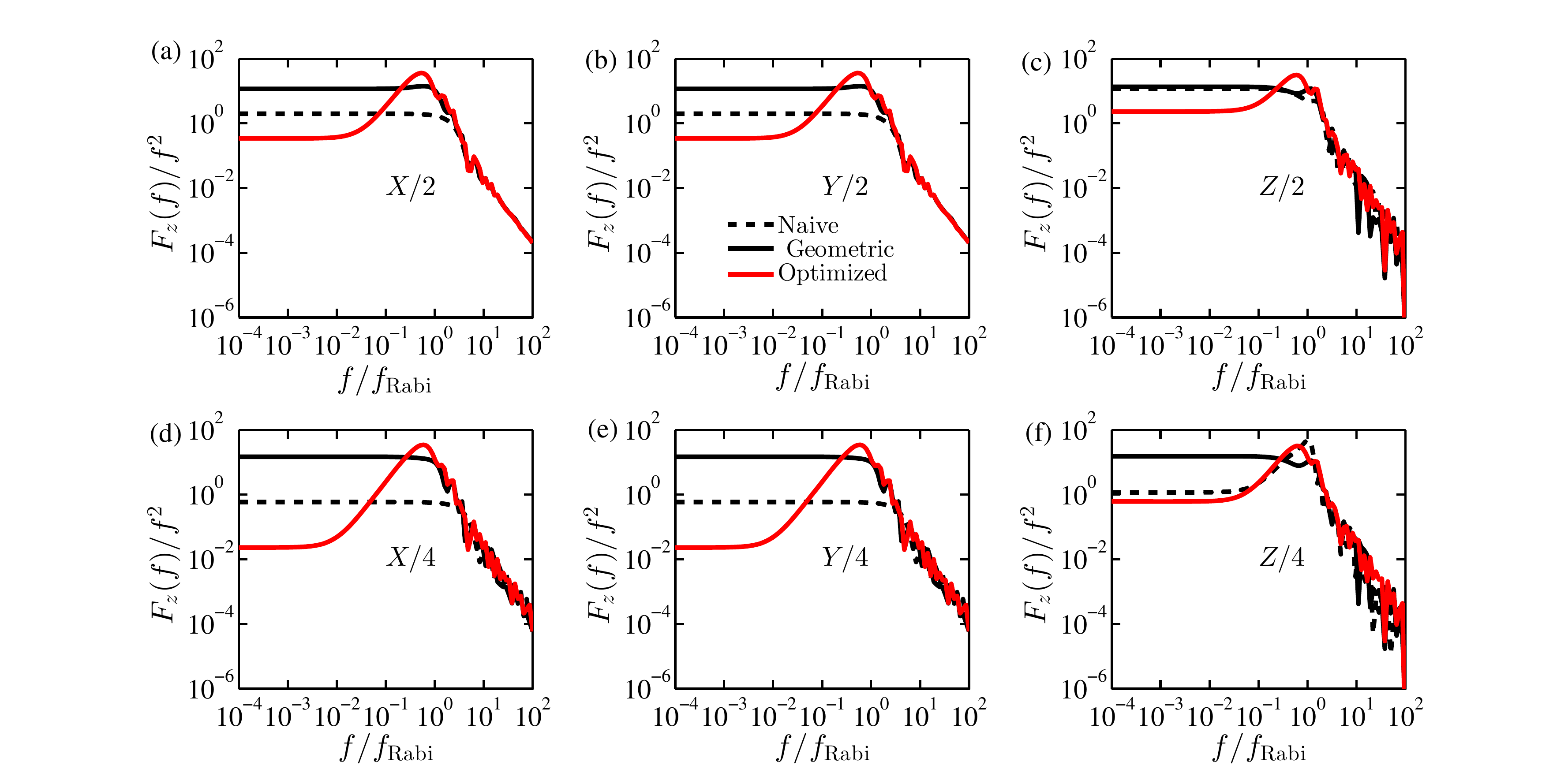}
	\caption{Filter function results. The noise is assumed to be with $1/f$ type, where the spectral density can be described as $S(f)=\frac{S_{0}}{f^{\alpha}}$. In the simulation, we take the parameter from the experimental results in Ref.~\cite{Yoneda.18}: $S_{0}=2.67\times 10^{6}\ \text{Hz}^{2}$, $\alpha=1.01$, $\Omega/2\pi=f_{\rm{Rabi}}=4\ \rm{MHz}$, the cutoff $f_{\rm{uv}}=320$ kHz.}
	\label{fig:filter}
\end{figure*}

The standard benchmarking results for the naive dynamical, the conventional geometric and the optimized geometric gates are: $\mathcal{F}_{d}=99.957 \%$, $\mathcal{F}_{g}=99.919 \%$ and $\mathcal{F}_{g,opt }=99.998 \%$. It is clear that the optimized geometric gate performs the best. With the standard randomized benchmarking results, one can further obtain the interleaved benchmarking results via $\mathcal{F}_{\text {in }}=1-(1-p_{\mathrm{in}} / p_{\mathrm{st}})/2$ \cite{Easwar.12,Chen.22}, where $p_{\mathrm{in}}$ and $p_{\mathrm{st}}$ denote the depolarizing parameters for the interleaved and standard benchmarking results. The depolarizing parameter is determined by $p=e^{-d}$. The interleaved benchmarking results are shown in Fig.~\ref{fig:rb2}, while the related gate fidelities are shown in Table.~\ref{table:rb}. One is clear that for all the cases, the optimized geometric gates outperform the naive dynamical and the conventional geometric ones.
\begin{table}
	\scalebox{0.83}{
		\begin{tabularx}{10.35cm}{ccccccc}
			\hline
			\hline
			Gates& $X/2$ \ \ \ & $X/4$ \ \ \ & $Y/2$  \ \ \ & $Y/4$  \ \ \ & $Z/2$  \ \ \  & $Z/4$  \\
			\hline
			\text { Naive } & 99.957\% & 99.953\% & 99.954\% & 99.960\% & 99.884\% & 99.969 \%
			\\ 	
			\text { Geometric } & 99.919\% & 99.878\% & 99.898\% & 99.868\% & 99.913\% & 99.905\%
			\\
			\text { Optimized } & 99.978\% & 99.975\% & 99.976\% & 99.974\% & 99.975\% & 99.977\%
			\\
			\hline
			\hline
	\end{tabularx}}
	\caption{Fidelity of the interleaved randomized benchmarking, drawn by fitting the data in Fig.~\ref{fig:rb2}. }
	\label{table:rb}
\end{table}

On the other hand, for silicon-based semiconductor quantum dot in experiments, the noise is time-dependent, and is correlated. Normally, the noise spectrum  exists to be the $1/f$ type, whose spectral density can be described as
\begin{equation}
\begin{aligned}
S(f)=\frac{S_{0}}{f^{\alpha}},
\end{aligned}
\label{sw}
\end{equation}
where $f$ is the frequency, $S_{0}$ denotes the noise amplitude, and the exponent $\alpha$ denotes how much the noise is correlated. For small $\alpha$, the noise is strongly correlated and is far away the static noise model. while for large $\alpha$, the noise is gradually closed to the static model. According to a recent experiment \cite{Yoneda.18}, we have $S_{0}=2.67\times 10^{6}\ \text{Hz}^{2}$, $\alpha=1.01$, which means the noise is highly correlated. Therefore, the performance of the optimized geometric gates under the realistic noise environments still needs to be verified.

The filter function \cite{Green.12,Green.13,Silva.14,Barnes.16} is an powerful tool to analytically calculate the infidelity for the time-dependent errors in the qubits. The calculated infidelity during the control operation can be expressed with respect to the specific noise spectrum and the control Hamiltonian. Here, we briefly introduce how to calculate the filter function. The key of the filter function is the so-called control matrix $\boldsymbol{R}(t) \equiv\left[\boldsymbol{R}_{x}(t), \boldsymbol{R}_{y}(t), \boldsymbol{R}_{z}(t)\right]^{\mathrm{T}}$, the components of which are
\begin{equation}
\begin{aligned}
\boldsymbol{R}_{j}(t) &\equiv\left[R_{j x}(t), R_{j y}(t), R_{j z}(t)\right],\\	
R_{j k}(t) &=\frac{\operatorname{Tr}\left[U_{c}^{\dagger}(t) \sigma_{j} U_{c}(t) \sigma_{k}\right]}{2},
\end{aligned}
\label{con}
\end{equation}
where $j, k \in\{x, y, z\}$. Here, $U_{c}(t)$ is the time-dependent evolution operator that is satisfied with the Schrödinger equation $i \dot{U}_{c}(t)=H(t) U_{c}(t)$. Then the Fourier transform of $R_{j k}(t)$ is 
\begin{equation}
\begin{aligned}
R_{i j}(f)=-i f \int_{0}^{T'} e^{i f t}  R_{i j}(t) d t,
\end{aligned}
\label{con2}
\end{equation}
where $T'$ is the duration of the total gate operation. For a given noise spectrum $S_{ij}(f)$, the fidelity is 
\begin{equation}
\begin{aligned}
\mathcal{F} \simeq 1-\frac{1}{2 \pi} \sum_{i, j, k=x, y, z} \int_{-\infty}^{\infty} \frac{d f}{f^{2}} S_{i j}(f) R_{j k}(f) R_{i k}^{*}(f).
\end{aligned}
\label{filterf}
\end{equation}
Since we only consider the off-resonance error, which appears in the $z$ axis in the control Hamiltonian, and the specific frequency domain in the experiment, Eq.~\ref{filterf} reduces to
\begin{equation}
\begin{aligned}
\mathcal{F}=1-\frac{1}{2 \pi} \int_{0}^{f_{\rm{uv}}} d f S_{z}(f) \frac{F_{z}(f)}{f^{2}}.
\end{aligned}
\label{filterf2}
\end{equation}
Here we have define the filter function as 
\begin{equation}
\begin{aligned}
F_{z}(f)=\sum_{k=x, y, z} R_{z k}(f) R_{z k}^{*}(f),
\end{aligned}
\label{filter}
\end{equation}and $f_{\rm{uv}}$ is the cutoff related to the experiment. From Eq.~\ref{filterf2}, it is clear that for a specific noise spectrum $S_{z}(f)$, the related filter function term $F_{z}(f)/f^{2}$ can be used to evaluate the infidelity of the operation, since it is proportional to the infidelity. The filter function results are shown in Fig.~\ref{fig:filter}. For all the cases, when the frequency is small ($f/f_{\rm{Rabi}}\leq10^{-1}$), the filter function with respect to the optimized geometric gate is the smallest. When the frequency is large enough ($10^{-1}\leq f/f_{\rm{Rabi}}\leq10^{0}$), the filter function with respect to the naive dynamical gate turns to be the smallest. When the frequency is too large, the filter function curves for all the gates are indistinguishable. On the other hand, the curve denoted as the conventional geometric gate always above the naive dynamical one. In short, the optimized geometric gate is robust against the low-frequency off-resonance noise, while the conventional geometric gate is helpless for the off-resonance noise.

By inserting Eq.~(\ref{sw}) into (\ref{filterf2}), one is able to further calculate the numerical fidelity under the $1/f$ noise spectrum. In our simulation, we have considered the related parameters from the experiment \cite{Yoneda.18}: the Rabi frequency $\Omega/2\pi=f_{\rm{Rabi}}=4\ \rm{MHz}$, the cutoff $f_{\rm{uv}}=320$ kHz (corresponding to $f/f_{\rm{Rabi}}=8\times10^{-1}$). The related calculated fidelities are shown in Table.~\ref{table:filter}. For all the cases, the optimized geometric gates still perform the best.

\begin{table}
	\scalebox{0.81}{
		\begin{tabularx}{10.3cm}{ccccccc}
			\hline
			\hline
			Gates& $X/2$ \ \ \ & $X/4$ \ \ \ & $Y/2$  \ \ \ & $Y/4$  \ \ \ & $Z/2$  \ \ \  & $Z/4$  \\
			\hline
			\text { Naive } & 99.615\% & 99.887\% & 99.615\% & 99.887\% & 97.693\% & 99.775\%  
			\\ 	
			\text { Geometric } & 97.779\% & 97.153\% & 97.759\% & 97.153\% & 97.374\% & 97.041\%
			\\
			\text { Optimized } &  99.934\% & 99.996\% & 99.934\% & 99.999\% & 99.550\% & 99.883\%
			\\
			\hline
			\hline
	\end{tabularx}}
	\caption{Fidelity results from the filter function, drawn by fitting the data in Fig.~\ref{fig:filter}.}
	\label{table:filter}
\end{table}

\section{Conclusions }\label{sec:conclusion}

In this work, we have proposed a general method to obtain the optimized geometric gate by using dynamical correction. We find that by selecting a new evolution path and properly inserting a $\pi$-pulse into the evolution path, the obtained optimized geometric gate is equivalent to the conventional one. To verify the robustness of the optimized geometric gates against the off-resonance error, we further perform randomized benchmarking and calculate the filter function under the realistic noise level from experiments. The results clearly show that the optimized geometric gates are superior to the conventional geometric gates and the naive dynamical gates without using the geometric phase. Our results indicate dynamical correction might be useful to improve the geometric gate.

\section*{ACKNOWLEDGMENTS}\label{sec:ack}
This work was supported by the National Natural Science Foundation of China (Grant No. 11905065, 62171144), and the Guangxi Science Foundation (Grant No. AD22035186, 2021GXNSFAA220011).

\begin{figure}
	\includegraphics[width=0.9\columnwidth]{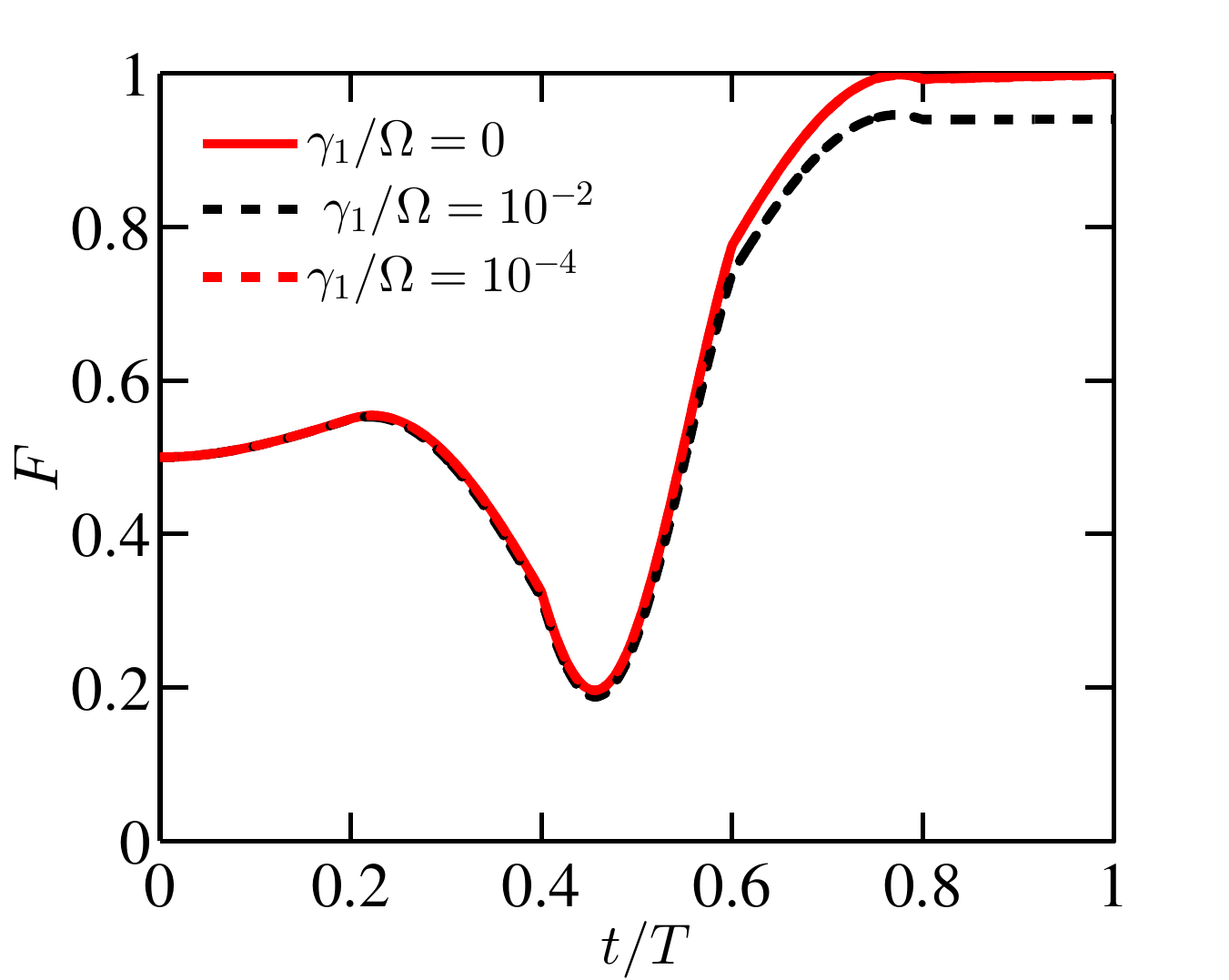}
	\caption{Decoherence effect due to the relaxation. The off-resonance error is $\delta=0.1$. When the relaxation rate is large, i.e., $\gamma_{1}/\Omega=10^{-2}$ (denoted as black dashed line), the fidelity curve would slightly deviate from the one without relaxation effect (denoted as red solid line). While the relaxation rate is small enough, i.e., $\gamma_{1}/\Omega=10^{-4}$ (denoted as red dashed line), the introduced infidelity by the relaxation can be safely neglected.}
	\label{fig:app1}
\end{figure}

\appendix

\section{Decoherence effect}\label{appx2}

Compared with the conventional geometric gate, the evolution time for the optimized one is longer (adding $\pi$ more pulse), which might cause more decoherence when considering the relaxation. To study this decoherence effect on the optimized geometric gate, we introduce the master equation
\cite{blais.07}
\begin{equation}
	\begin{aligned}
	\dot{\rho}=-i[H_{c}'(t), \rho]+ \gamma_{1} \mathcal{D}\left[\sigma_{-}\right] \rho+ \frac{\gamma_{\varphi}}{2} \mathcal{D}\left[\sigma_{z}\right] \rho,	
	\label{eq:master}
\end{aligned}
\end{equation}
where $\mathcal{D}[\hat{L}] \rho=\left(2 L \rho L^{\dagger}-L^{\dagger} L \rho-\rho L^{\dagger} L\right) / 2$.
Here, $\rho$ describes the density matrix, $\sigma_{-}=|0\rangle \langle 1|$ and $\sigma_{z}=|0\rangle \langle 0|-|1\rangle \langle 1|$. $\gamma_{1}$ and $\gamma_{\varphi}$ denote the relaxation and pure dephasing rate, respectively. In Fig.~\ref{fig:app1}, we show the fidelity of the $X/2$ rotation versus the evolution time, considering different relaxation rates. Here we take the off-resonance error to be $\delta=0.1$ which contributes to the dephasing. Thus we set $\gamma_{\varphi}=0$. As we can see that the large relaxation rate would result in considerable infidelity. While the relaxation rate is small enough with $\gamma_{1}/\Omega=10^{-4}$ (as seen the red dashed line), the infidelity can be as low as $\sim 10^{-4}$. For the spin qubits in silicon, the relaxation time can be as long as 9 s \cite{Ciriano.21}. Considering the normal Rabi frequency $\Omega/2\pi=500$ kHz \cite{Huang.19}, the corresponding relaxation rate can be $\gamma_{1}/\Omega\sim 10^{-8}$. In this way, it is safe to neglect the relaxation effect.

\section{Geometric gates with two inserted $\pi$-pulse}\label{appx}

Here, we introduce the dynamically corrected geometric gates in Ref.~\cite{Li.21}, where two  $\pi$-pulse are inserted into the middle point of the evolution compared to the conventional orange-slice-shaped geometric gates. Specifically, $\Omega(t)$ and $\varphi(t)$ are divided into seven pieces, where they satisfy
\begin{equation}
	\begin{aligned}
		\int_{0}^{T_1}\Omega(t)dt&= \theta, \quad \left\{\varphi(t)\equiv\phi-\frac{\pi}{2},t\in\left[0,T_1\right]\right\}\\
		\int_{T_1}^{T_2}\Omega(t)dt&= \frac{\pi}{2}, \quad \left\{\varphi(t)\equiv\phi+\gamma+\frac{\pi}{2},t\in\left[T_1,T_2\right]\right\}\\
		\int_{T_2}^{T_3}\Omega(t)dt&= \pi, \quad \left\{\varphi(t)\equiv\phi+\gamma+\pi,t\in\left[T_2,T_3\right]\right\}\\
		\int_{T_3}^{T_4}\Omega(t)dt&= \frac{\pi}{2}, \quad \left\{\varphi(t)\equiv\phi+\gamma+\frac{\pi}{2},t\in\left[T_3,T_4\right]\right\}\\
		\int_{T_4}^{T_5}\Omega(t)dt&= \frac{\pi}{2}, \quad \left\{\varphi(t)\equiv\phi-\frac{\pi}{2},t\in\left[T_4,T_5\right]\right\}\\
		\int_{T_5}^{T_6}\Omega(t)dt&= \pi, \quad \left\{\varphi(t)\equiv\phi,t\in\left[T_5,T_6\right]\right\}\\
		\int_{T_6}^{T}\Omega(t)dt&= \frac{\pi}{2}-\theta, \quad \left\{\varphi(t)\equiv\phi-\frac{\pi}{2},t\in\left[T_6,T\right]\right\}.
	\end{aligned}
	\label{eq:seg3}
\end{equation}
For an arbitrary $x$-axis rotation by an angle $\chi$, the corresponding fidelity can be expanded as 
\begin{equation}
	\begin{aligned}
		\mathcal{F}_{2\pi}^{\delta}(\hat{x},\chi)&\simeq 1+\cos\frac{\gamma^{2}}{4}(\cos\frac{\gamma}{2}-2\sin\frac{\gamma}{2}-3)\delta^{2}.
	\end{aligned}
	\label{eq:app1}
\end{equation}	
One can easily see that within the range $-\pi\leq \chi\leq \pi$, $\mathcal{F}_{2\pi}^{\delta}(\hat{x},\chi)\leq \mathcal{F}_{d}^{\delta}(\hat{x},\chi), \mathcal{F}_{g}^{\delta}(\hat{x},\chi), \mathcal{F}_{g,opt}^{\delta}(\hat{x},\chi)$. Therefore, this protocol is unable to improve the off-resonance error.


%

\end{document}